\documentclass[conference]{IEEEtran}
\IEEEoverridecommandlockouts
\usepackage{cite}
\usepackage{amsmath,amssymb,amsfonts}
\usepackage{algorithmic}
\usepackage{graphicx}
\usepackage{textcomp}
\usepackage{xcolor}
\def\BibTeX{{\rm B\kern-.05em{\sc i\kern-.025em b}\kern-.08em
    T\kern-.1667em\lower.7ex\hbox{E}\kern-.125emX}}

\usepackage{times}
\usepackage{soul}
\usepackage{url}
\usepackage{booktabs}
\usepackage{algorithm}
\usepackage{algorithmic}
\usepackage[switch]{lineno}
\usepackage{paralist}
\usepackage{multirow}
\usepackage{caption}
\usepackage{subcaption}

\begin{document}

\title{Prepared for the Unknown: Adapting AIOps Capacity Forecasting Models to Data Changes\\
}
\author{
  \IEEEauthorblockN{Lorena Poenaru-Olaru\IEEEauthorrefmark{1}, Wouter van 't Hof\IEEEauthorrefmark{2}, Adrian Sta\'ndo\IEEEauthorrefmark{3}, Arkadiusz P. Trawi\'nski\IEEEauthorrefmark{3}, Eileen Kapel\IEEEauthorrefmark{2},\\ Jan S. Rellermeyer\IEEEauthorrefmark{4}, Luis Cruz\IEEEauthorrefmark{1}, Arie van Deursen\IEEEauthorrefmark{1}}
  \IEEEauthorblockA{\IEEEauthorrefmark{1}Software Engineering Research Group, \textit{Delft University of Technology}, Delft, The Netherlands \\
  \{L.Poenaru-Olaru, l.cruz, Arie.vanDeursen\}@tudelft.nl}
  \IEEEauthorblockA{\IEEEauthorrefmark{2}Engineering \& Reliability, \textit{ING Netherlands}, Amsterdam, The Netherlands \\
  \{wouter.van.t.hof, Eileen.Kapel\}@ing.com}
  \IEEEauthorblockA{\IEEEauthorrefmark{3}\textit{ING Hubs Poland}, Warsaw, Poland \\
  \{Adrian.J.Stando, Arkadiusz.Trawinski\}@gmail.com}
  \IEEEauthorblockA{\IEEEauthorrefmark{4}Dependable and Scalable Software Systems, \textit{Leibniz University Hannover}, Hannover, Germany \\
  rellermeyer@vss.uni-hannover.de}
}
\maketitle

\begin{abstract}
Capacity management is critical for software organizations to allocate resources effectively and meet operational demands. An important step in capacity management is predicting future resource needs often relies on data-driven analytics and machine learning (ML) forecasting models, which require frequent retraining to stay relevant as data evolves. Continuously retraining the forecasting models can be expensive and difficult to scale, posing a challenge for engineering teams tasked with balancing accuracy and efficiency. Retraining only when the data changes appears to be a more computationally efficient alternative, but its impact on accuracy requires further investigation. In this work, we investigate the effects of retraining capacity forecasting models for time series based on detected changes in the data compared to periodic retraining. Our results show that drift-based retraining achieves comparable forecasting accuracy to periodic retraining in most cases, making it a cost-effective strategy. However, in cases where data is changing rapidly, periodic retraining is still preferred to maximize the forecasting accuracy. These findings offer actionable insights for software teams to enhance forecasting systems, reducing retraining overhead while maintaining robust performance.

\end{abstract}


\section{Introduction}
The term capacity management refers to ensuring that an IT service has sufficient infrastructure and resources to meet the current or future demand. Although capacity management is crucial to ensure efficient and effective service delivery, this process used to be carried on manually by continuously collecting and analyzing data~\cite{Wang2022ANT}. Manual techniques to predict the capacity requirements become difficult to scale as the capacity management data sources increase, and it is significantly time-consuming for the engineers in charge.

To automate the capacity management for machine utilization, like CPU and memory, companies have started employing forecasting AIOps models, which predict the resource demand in a timely fashion. This is particularly relevant for our industry partner, ING (International Netherlands Group) Bank, where operational engineers must monitor numerous time series to ensure sufficient resources are allocated for its large-scale online operations, supported by thousands of machines with varying resource demands. As our case study, we use a capacity forecasting model developed within ING by data scientists. This forecasting model is trained on historical time series operational data related to CPU and memory utilization and predicts the number of necessary resources for the upcoming two weeks. By leveraging this model, ING significantly reduces the manual effort required to analyze historical data and predict future demand, while also minimizing the risks of insufficient or excessive resource allocation.


Real-world operational data usually has an evolving character, meaning that it constantly changes over time as it is influenced by shifts in customer behavior or infrastructure.~\cite{datasplittingdecisions, mypaperanomalydetection, towardsaconsistentinterpretation}. For instance, a common maintenance task, such as software or hardware updates, is known to induce fundamental changes in operational data~\cite{nodefailurepredictionconceptdrift}. Although changes in data, commonly referred to as concept drift~\cite{conceptdriftadaptation}, are known to impact the performance and reliability of AIOps forecasting models over time~\cite{Wang2022ANT, Isaac2023QBSDQS}, their specific impact on forecasting accuracy has yet to be systematically quantified. Therefore, one of the contributions of the work is an empirical study of forecasting models' performance degradation due to data changes over a predefined period of time in a real-world industry setting. Our approach can be systematically used for other time series forecasting applications within various other industries.

One solution to overcome the issue of continuous data changes is periodic retraining~\cite{datasplittingdecisions, towardsaconsistentinterpretation, modelmaturity, mypaperanomalydetection, Isaac2023QBSDQS}. Industry researchers~\cite{Isaac2023QBSDQS} have highlighted concerns about the scalability of periodic retraining due to the computational costs of managing hundreds of time series. Therefore, our study further contributes by analyzing the benefits of continuous retraining.


Although previous work has showcased the potential of using data change (drift) detection techniques with AIOps anomaly detection models~\cite{mypaperanomalydetection}, to the best of our knowledge, this is the first study that investigates the benefits of retraining the capacity forecasting model within a large real-world software organization. Therefore, the third contribution of our work is investigating the benefits of retraining the capacity forecasting model used within the chosen case company only when data have changed, vs. retraining periodically. We are further presenting the implications of employing a drift detector in a practical setting, as well as retraining costs reduction, and model performance changes.


\section{Motivational Use Case}


In order to improve capacity management, ING is using a forecasting model which aims to predict the necessary resources for a predefined period of time. This model is currently applied to an experimental size of 16 time series (CPU and memory utilization time series) extracted from eight machines as a proof of concept, which will be used for this study. It is expected that in the future this solution will be scaled up to a significantly higher number when it is made generally available as a service in the case company.

To ensure that the forecasting model is continuously updated and aligned with potential changes in the data model, retraining is required. The model is retrained according to a retraining scheduler, and each time series has its own forecasting model. The retraining scheduler is currently programmed to retrain the forecasting model on a periodic basis every month. Although retraining more frequently, such as weekly or biweekly, could be beneficial, the responsible data scientist encountered scalability issues in terms of the computation required to retrain the model for all time series. For instance, even if the duration of retraining, including hyperparameter optimization and training the model with new data, is relatively short per time series instance when being scaled to a high number of time series, it becomes substantial. Therefore, although retraining more frequently, such as weekly or bi-weekly, could be beneficial, the models would require too much time to be updated, given the amount of data that we continuously collect. Besides this, regarding model scalability, applying our forecasting model to even more time series will further considerably increase the retraining. Furthermore, the process of redeploying a model after retraining in big organizations also implies plenty of validation. For this reason, we considered that monthly retraining is the best choice for our particular use case. 

Another benefit of reducing the time spent on continuous retraining is the potential improvement in forecasting model performance. Currently, the hyperparameter tuning phase during retraining is limited to a small set of hyperparameters due to time constraints. By reducing the number of models that need retraining, the server can allocate more time to explore additional configurations. With fewer retrainings required overall, the server can free up resources that would otherwise be spent on retraining all the models for different time series. This has the potential to enhance the model's accuracy and robustness. Therefore, transitioning from periodic retraining to retraining only when necessary due to data changes can not only reduce the time needed for updates but also potentially improve model performance by allowing more hyperparameter configurations to be explored. 



\section{Research Questions}

Although retraining only when data changes is a promising approach, we have to investigate the implications of doing so in terms of model performance and reduction in retraining time. For this reason, we begin this study with an understanding of whether we observe a severe model degradation over time when the model is never updated. This would allow us to understand the evolution of the model's performance over time. We further compare two scenarios: retraining monthly, which is our current practice, and retraining only when data changes are signaled by a drift detector. In this study, we aim to answer the following research questions:
\begin{compactenum}
\item[\textbf{RQ1:}] What is the evolution of the performance of the forecasting model over time?
\item[\textbf{RQ2:}] What is the difference between retraining based on drift detection and periodic retraining in terms of forecasting model accuracy and retraining frequency? 
\end{compactenum}

\section{Related Work}
This section begins with an overview of the AIOps domain and the challenges faced by AIOps models due to changes in data over time, particularly in applications dealing with time series data. We then introduce drift detection techniques tailored for time series, providing a detailed explanation of the drift detector selected for this study.
\subsection{AIOps}
AIOps refers to employing Artificial Intelligence (AI) and Machine Learning (ML) techniques to solve complex DevOps challenges~\cite{salesforceai}. The adoption of AIOps has the potential to decrease operational costs through automation, enhance engineering productivity, and ensure high-quality services by predicting possible failures or system anomalies~\cite{aiopschallenges}. A recent survey by SalesforceAI classifies AIOps applications into four categories: failure prediction, incident detection, root-cause analysis, and automated actions. Examples of failure prediction AIOps applications are predicting hard disk drives (HDDs) in large data centers~\cite{diskfailureprevwork, diskfailureprevwork1, datasplittingdecisions, towardsaconsistentinterpretation, modelmaturity} for hardware management purposes, node failure prediction in large-scale cloud service systems~\cite{nodefailure1}, and job failure prediction~\cite{googletraceprediction1, datasplittingdecisions, towardsaconsistentinterpretation, modelmaturity}. In their work, Kapel et al.~\cite{eileenpaper} present incident detection techniques that were successfully applied in industry, namely~\cite{incident1} and~\cite{incident2}. Root-cause analysis applications aim to reduce the time required to identify the cause of an incident by identifying abnormal patterns in Key Performance Indicators (KPIs)~\cite{rootcauseanalysis1, rootcauseanalysis2}. The automated actions category aims to automate tasks performed manually by operational engineers, such as automated remediation, auto-scaling, and resource management~\cite{salesforceai}. Our studied use case, capacity forecasting, is part of the automated actions - resource management category.

\subsection{Adapting AIOps Models to Data Changes}

Given that changes in the operational data impact the performance of AIOps solutions, plenty of attention has been paid to adapting different AIOps models to changes in the data over time. Lyu et al.~\cite{datasplittingdecisions, towardsaconsistentinterpretation} have shown that failure prediction AIOps models require periodic updates to preserve the accuracy over time. Moreover, in~\cite{datasplittingdecisions} the authors claim that a higher updating frequency usually leads to better performance. Anomaly detection AIOps models have also been investigated in terms of adaptation to concept drift based on retraining~\cite{ericsonanomalydetection, mypaperanomalydetection}. Furthermore, besides periodic retraining, retraining AIOps models based on concept drift detection has been proposed~\cite{myshortpaper, ericsonanomalydetection, mypaperanomalydetection}. However, to the best of our knowledge, no study analyzes the suitability of retraining based on drift detection when it comes to capacity forecasting models.

\subsection{Drift Detection for Time Series}
In their exhaustive survey about concept drift detectors, Bayram et al.~\cite{BAYRAM2022108632} highlighted that, while plenty of drift detectors were developed for classification problems, few drift detectors exist for time series. The reason for this discrepancy is the lack of availability of relevant time series datasets. In this subsection, we provide a general overview of the existing concept drift detection techniques for time series, and we explain the functionality of the drift detector we employ in this study.

\subsubsection{General Overview Drift Detection for Time Series}
The comprehensive study of Bayram et al.~\cite{BAYRAM2022108632} reveals that while there has been considerable research on drift detectors for classification problems, there has been significantly less focus on those suitable for time series data. One reason for this is the lack of availability of open-source data on which the drift detection performance could be evaluated.

The most popular drift detector for time series is Feature Extraction Drift Detection (FEDD)~\cite{fedd}. This drift detector was previously applied to real-world stock market data provided by Yahoo Finance~\cite {feddPSO}. Another proposed drift detector for time series data is Entropy-Based Time Domain Feature Extraction (ETFE)~\cite{efte}, which was only evaluated on synthetic data. Both drift detectors require a feature extraction phase in which features are computed from each time series. However, ETFE is more computationally intensive since the features it requires are based on time series decomposition, while FEDD does not perform additional transformations of the time series. Due to scalability issues while continuously computing the features required for drift detection for each time series in real-time, ETFE is deemed unsuitable since reducing computation is a key requirement for the forecasting application. Given that efficiency is a major requirement and FEDD is less computationally intensive, we decided to employ FEDD in our experiments. Furthermore, due to internal output extraction policies, we exclude drift detectors that continuously measure the performance of the forecasting model over time to identify significant drops, such as DDM~\cite{ddm}, EDDM~\cite{eddm}, or ADWIN~\cite{adwin} and solely focus on detectors that identify drift from the time series data. 

\subsubsection{FEDD}

FEDD is a drift detection technique designed for time series data that identifies drift based on features extracted from the time series itself. Thus, FEDD identifies drift solely from the changes that can be observed in the time series, without considering how these changes impact the performance of a forecasting model that uses the data.

The chosen drift detector consists of two main components: the feature extraction module and the drift detection module. The feature extraction module calculates time series features that will be used to detect changes. FEDD takes into account six linear features (variance, autocorrelation, skewness coefficient, partial autocorrelation, turning point rate, and kurtosis coefficient) and two non-linear features (mutual information and bicorrelation). The drift detection component computes the similarity between two feature vectors corresponding to two predefined time series windows using the cosine distance. Thereafter, the exponentially weighted moving average (EWMA) is employed to understand whether the similarity is significant and the drift needs to be signaled.

In terms of practical functionality, an initial reference time series window needs to be defined, and the initial feature vector needs to be extracted from this window. FEDD works in an online manner, namely, the reference window is constantly shifted by one sample to define the current time series window and to extract the current feature vector. The similarity between the two feature vectors, initial and current, is computed and stored in an array from which EWMA identifies whether there was a significant change in similarity. Once a drift is identified, the reference window needs to be redefined. To avoid a high number of false alarms, the reference window is shifted by a predefined number of samples. In a practical setting, this shift implies that the drift detectors go to a cool-down period in which they will be inactive until enough samples are collected to reinitialize the reference window. Once the new reference window is defined, the drift detector restarts and is able to monitor data changes in the time series again.




\section{Methodology}
In this section, we present our experimental setup in terms of the datasets used, a detailed explanation of the forecasting models employed in this study, the evaluation metric used to assess the performance of the forecasting model, and the performance of retraining based on drift detection.
\subsection{Datasets}
We use proprietary datasets consisting of 16 time series, with half related to memory utilization and the other half to CPU utilization. These datasets were collected from real-world ING servers over a period of approximately nine months (from February until November 2023). The data was originally collected at a minute-level granularity, but to ensure confidentiality, we aggregate it to one hour by taking the mean of the samples. Although we only experimented with 16 time series, these time series are representative of data extracted from real-world financial infrastructure.

\subsection{Forecasting Model Description}

We employ the forecasting model that ING  is currently using it to predict resource capacity in terms of CPU and memory utilization. In this subsection, we briefly present the forecasting model's functionality and the features used to train it.

\subsubsection{Forecasting Functionality}

The forecasting model is designed to predict CPU and memory utilization for a two-week horizon. It is applied weekly to forecast the upcoming two weeks. As a result, the forecast for the first week is expected to be the most accurate, while the prediction for the second week is more estimative but still essential for effective resource management. To simplify the evaluation, we consider that both weeks should be predicted accurately when computing the evaluation metric of the forecasting model. The forecasting model is initially trained on approximately one-third of each available time series, leaving the rest for testing and experimentation of the effect of retraining.

In order to predict the next sequence, different features are extracted from the time series. A detailed overview of the time series features is presented in the following section. To ensure that the forecasting model can predict the future using information from the present, these features are also calculated taking into account a forecasting horizon. For instance, the sample in the time series collected at the current moment (the true label) corresponds to the features computed two weeks ago.

For each time series, a LightGBM regressor is trained using the first eight weeks of data and validated in the following two weeks. The best parameters for the LightGBM model are determined using the validation set. The remaining weeks are employed to evaluate the effects of retraining the model based on drift detection vs. periodically. 

\subsubsection{Features} The forecasting model employs different categories of features for time series prediction, namely time, lag, rolling window, and Prophet features. 

The \textit{time features} of each time series sample refer to attributes related to the specific timestamp of the sample. These can include binary features, such as whether the sample was collected during the weekend or at the start/end of the month, as well as non-binary features like the month, quarter, or day of the month when the sample was recorded. These features are derived solely from the date of collection, without considering the actual value of the time series.
The \textit{lag features} are features computed by taking into account values of the time series in the past. These features work under the assumption that the current value of the time series is influenced by its past values, and the past values are meaningful in predicting future values. These features are computed based on a predefined forecasting horizon. 

The \textit{rolling window features} are also computed by taking into account past values of the time series. These features require a predefined time window, referring to the window required to calculate the features.

The \textit{Prophet features} are commonly used features in machine learning applications for forecasting at scale~\cite{prophet}. These features are useful when time series present strong seasonal patterns or trends. Prophet features are generated by decomposing each time series into three main components, namely seasonality, holidays, and trend. In our situation, Prophet features were added to the model since the resources are used differently according to the specific time of the week, month, or year.






\subsection{Retraining based on Drift Detection}
In this subsection, we present how we integrated the drift detector with our forecasting model setup. We highlight some challenges that we encountered in employing FEDD in a real-world machine learning application and propose a machine learning system design that includes drift detection-based retraining.
\subsubsection{Integration Challenges} While designing the architecture for the forecasting model to be retrained based on drift detection, we encountered two primary challenges: handling missing data and determining the optimal retraining timing.

\par{\textbf{Missing Data.}} 
In real-world scenarios, gaps in time series are inevitable due to system failures or updates that interrupt data collection. This does not affect our existing forecasting model, as we use a forecasting algorithm that is resilient to missing data. However, during our experiments, we observed that missing data presents challenges when detecting data drift using FEDD, which had not been an issue in previous evaluations on open-source continuous data. The problem arises because FEDD relies on time series features to detect drift, and these features can only be computed if the time series is continuous. To address this in our drift detection process, we treat the missing values as absent and reconstruct a continuous time series using the available data until a specified point.

\par{\textbf{Retraining Moment.}} An approach recommended in literature to overcome the impact of data changes in financial time series is presented by Cavalcante et al.~\cite{usingtsdriftdetectioninrealapplications}. The authors propose to perform online retraining, namely learning with every new upcoming sample, once the drift is detected, until the prediction error drops. However, this approach is not feasible for our use case since we use a batch-learning approach, namely, we retrain the model once a specific size of samples (e.g., one week of new samples) is collected. For this use case, switching to online learning is not feasible. The reason for this is that our model learns from the features derived from the time series instead of the time series itself. Some time series features are calculated based on a predefined window of samples that are taken into account. An online learning approach would imply that the features have to be continuously computed and updated with every new sample, which is computationally intensive given the number of time series to which this forecasting model will be applied. Furthermore, it has been shown that online learning is impacted by the irregularity of data caused by missing values~\cite{missingdataperspective}. The fact that our time series contains a significant amount of missing samples is another reason why we did not include an online learning approach in our solution to handle drift.


\subsubsection{Proposed Solution} To overcome the two encountered challenges, missing data and retraining moment, in this subsection, we describe our proposed solution in terms of drift detection integration.

\begin{figure}
    \centering
    \includegraphics[width=0.5\textwidth]{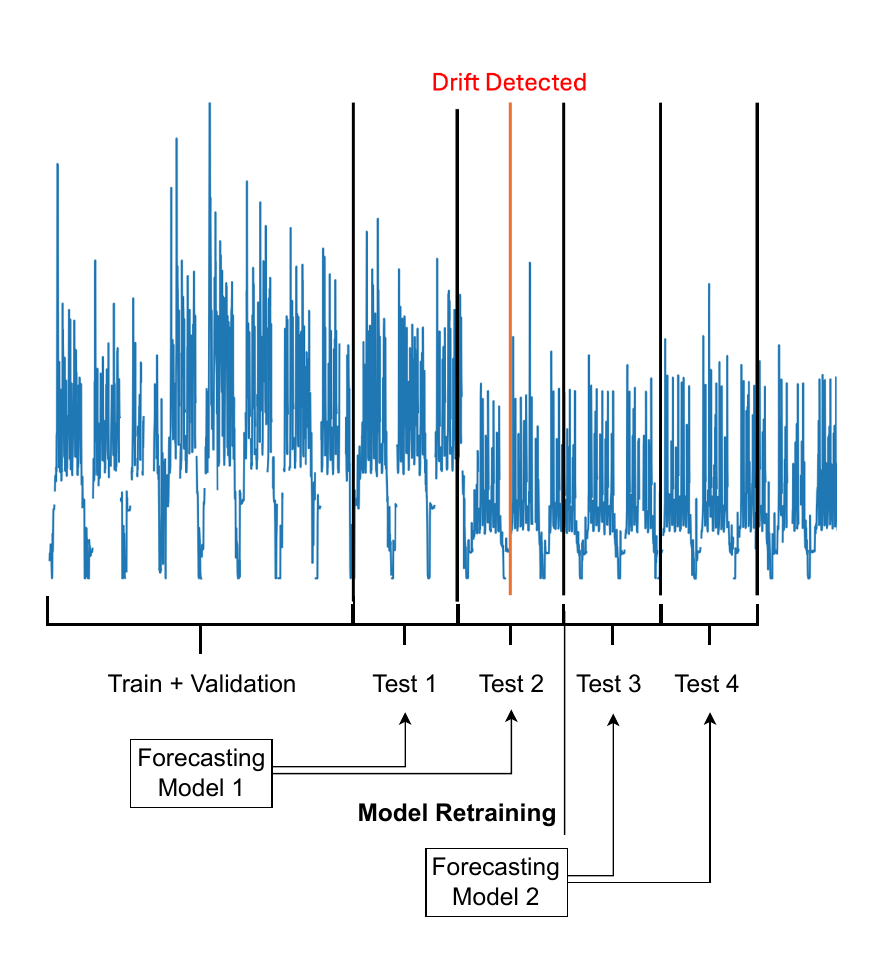}
    \caption{Retraining the forecasting model based on drift detection.}
    \label{figure:retraining_based_on_drift_detection}
    \vspace{-1em}
\end{figure}

In Fig.~\ref{figure:retraining_based_on_drift_detection}, we depict an example of a real-world time series from our datasets to explain how we propose to retrain the forecasting model based on drift detection. We start with an existing train and validation set that is used to train and tune the forecasting model. The model is further employed to predict the following two weeks as aforementioned. At the end of each week, we verify whether there was any drift signaled by the drift detector. If no drift is detected (e.g., the situation for Test 1), then the model is not updated. If drift is detected at the end of that week (e.g., the situation for Test 2), the model is retrained with all the available data at the end of the week where drift was identified and redeployed into production. Thus, the new model is used to obtain predictions for the following weeks.

\begin{figure*}
    \centering
    \includegraphics[width=0.75\textwidth]{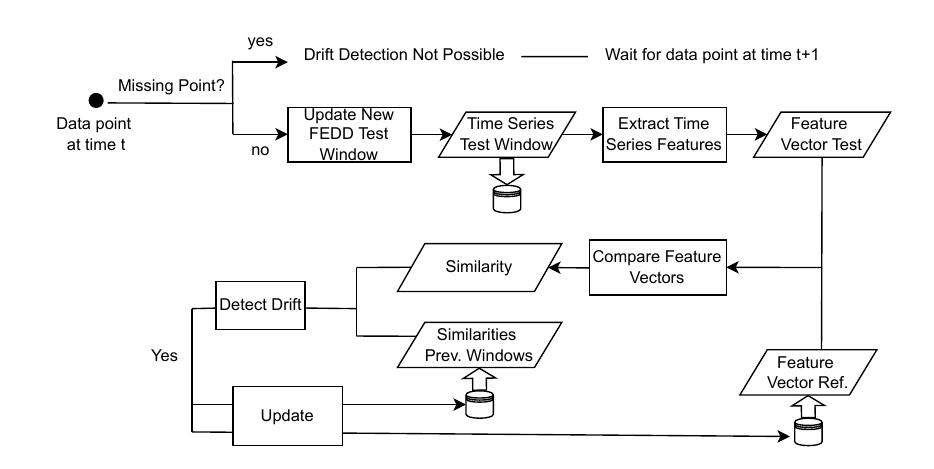}
    \caption{Drift Detection Block components and functionality.}
    \label{figure:drift_detection_block}
    \vspace{-1em}
\end{figure*}

In Fig.~\ref{figure:drift_detection_block}, we depict the workflow of the drift detector. From its original implementation, FEDD identifies drift in a real-time manner, namely that with every new sample of a time series, FEDD evaluates where drift has occurred. To integrate such functionality in a real-world case, taking into account the aforementioned limitations of this drift detection technique, at each point in time when the data should be collected, we determine whether the time series sample is missing or not. If there is a missing sample, then drift detection is not possible, and we have to wait until the next timestamp to collect data. In case of missing data, we do not perform interpolation in the time series. We solely ensure the continuity of the time series by concatenating the following non-missing sample to the current time series and therefore perform drift detection. If the sample is not missing, then we have to determine whether a drift was detected. In our setting, due to seasonality reasons, the reference window is set to eight weeks, and the current window is set to two weeks, the equivalent of a testing window. The next steps are extracting features out of the reference window and current window, respectively, storing the similarities in an array, and employing EWMA to identify whether there was a drift or not in the current testing window. Once a drift is detected, the reference window needs to be changed, and its corresponding feature vector has to be recalculated. The reference window has a fixed length, and it is changed every time a drift is identified by shifting the time window considered for the reference data. By doing so, we can avoid the drift detector signaling false alarms since it still contains samples from before the change in data is detected. 

One major advantage of our proposed drift detection block is the fact that not all time series require to be stored in order to detect drift, making FEDD an ideal solution when encountering scalability issues. As illustrated in Fig.~\ref{figure:drift_detection_block}, we only need to store the current evaluated time series window, the feature vector corresponding to the reference data, and the similarities array that gets continuously updated over time. Therefore, our proposed pipeline takes into account the minimization of storage space when performing drift detection and can be further employed in other time series forecasting applications where scalability is important. Furthermore, our solution is not specific to our case study, but it can be applied to any application working with time series data, such as other forecasting applications or even time series anomaly detection. Due to privacy issues, sharing the data and code is not possible. However, to encourage reproducibility, we created an open-source repository where practitioners can find a similar setting of employing FEDD on a sample time series dataset\footnote{\url{https://github.com/LorenaPoenaru/data_monitoring}}.

\subsection{Evaluation Metrics}
In this subsection, we present the evaluation metrics we employ in our study. The first part focuses on the evaluation metric used to assess the performance of the forecasting model. The second part presents how we assess the benefits of using drift detection-based retraining compared with our current monthly retraining technique.
\subsubsection{Forecasting Model Evaluation Metric}
To evaluate the performance of our forecasting models, we employ the mean absolute scaled error (MASE)~\cite{masemetric}. This metric measures the error made by a forecasting model compared to a naive forecasting approach. The formula used to compute MASE is the following:

\begin{equation} 
\text{MASE} = \frac{1}{n} \sum_{t=1}^{n} \left| \frac{Y_t - \hat{Y}_t}{\frac{1}{n-1} \sum_{i=2}^{n} |Y_i - Y_{i-1}|} \right| 
\end{equation}

where \( n \) is the number of data points in the time series, \( Y_t \) is the actual value at time \( t \), \( \hat{Y}_t \) is the predicted value at time \( t \) and \( Y_{t-1} \) is the previous value of the time series


Since it is an error metric, a lower MASE is desired when creating a forecasting model. This metric is chosen to evaluate our capacity forecasting model since it is robust to outliers that can occur in capacity forecasting due to periodic fluctuations in the systems' workloads~\cite{forecastevaluation}.

\subsubsection{Evaluation Metrics for Assessing Drift Detection vs Monthly Retraining}

In this section, we are presenting how we measure and assess the difference between retraining based on drift detection and our currently implemented scenario, monthly retraining. We compare the two retraining techniques from two perspectives: the \textit{model's performance} and the \textit{cost reduction}.

In terms of model performance, the first assessment metric that we calculate is the MASE improvement percentage of retraining based on drift detection (FEDD) over monthly retraining. The MASE improvement percentage can be either negative, meaning that the model retrained based on FEDD achieved a lower performance than the model that was retrained monthly, or positive, showing that the model retrained based on drift detection achieved a higher performance than the one retrained monthly. We calculate the MASE improvement percentage using the formula~\ref{erc} depicted below. Since MASE is an error-based metric, lower MASE values correspond to a better model. A negative MASE improvement suggests that the periodically retrained model obtained better predictions than the one retrained based on FEDD. In this situation, MASE\_FEDD and MASE\_Periodic are calculated by averaging the MASE over all time series batches used to test the forecasting model.

\begin{equation}\label{erc}
\mathit{MASE\_Impr.} = \frac{\mathit{MASE\_Periodic}-\mathit{MASE\_FEDD}}{\mathit{MASE\_Periodic}}*100
\end{equation}

In terms of cost reduction, we analyze the \textit{retraining savings} while retraining based on drift detection vs. monthly retraining. We define retraining savings (RS) in terms of to which we reduce the number of times the model is retrained by employing a drift detection retraining approach vs. a periodic retraining approach. RS are calculated using the formula~\ref{rs} shown below:

\begin{equation}\label{rs}
\mathit{RS} = \frac{\mathit{\#Retrainings\_Periodic}-\mathit{\#Retrainings\_FEDD}}{\mathit{\#Retrainings\_Periodic}}*100
\end{equation}

\section{Experimental Results}
In this section, we present the results obtained during the experimentation. Each subsection corresponds to the results of the experiments designed to answer each research question. Statistical significance of performance differences between settings was assessed using the Wilcoxon signed-rank test.

\subsection{Forecasting Performance over Time Analysis}


With this experiment, we analyze the evolution of the forecasting model's performance over time for all 16 analyzed time series, eight corresponding to CPU utilization and eight corresponding to memory utilization. We aim to answer RQ1 by understanding whether we can observe drops in the performance of the forecasting model over time, as also analyzed in previous AIOps applications for failure prediction~\cite{datasplittingdecisions, towardsaconsistentinterpretation, modelmaturity}.

\begin{figure*}
\centering
    \includegraphics[width=1.0\textwidth]{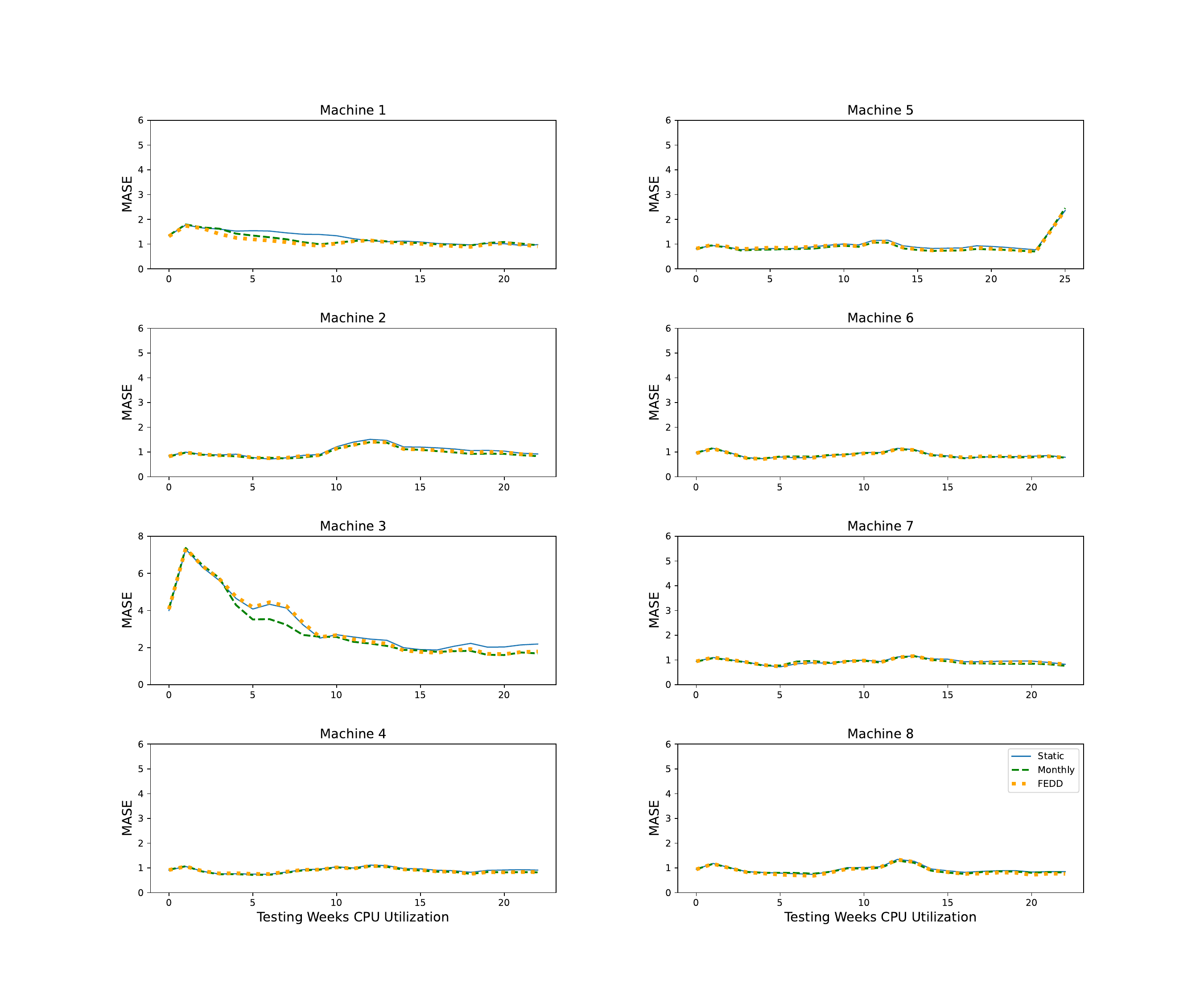}
    \caption{CPU Utilization}
    \label{figure:rq1_static_error_over_time_CPU}
    \vspace{-1em}
\end{figure*}

\begin{figure*}
    \centering
    \includegraphics[width=1.0\textwidth]{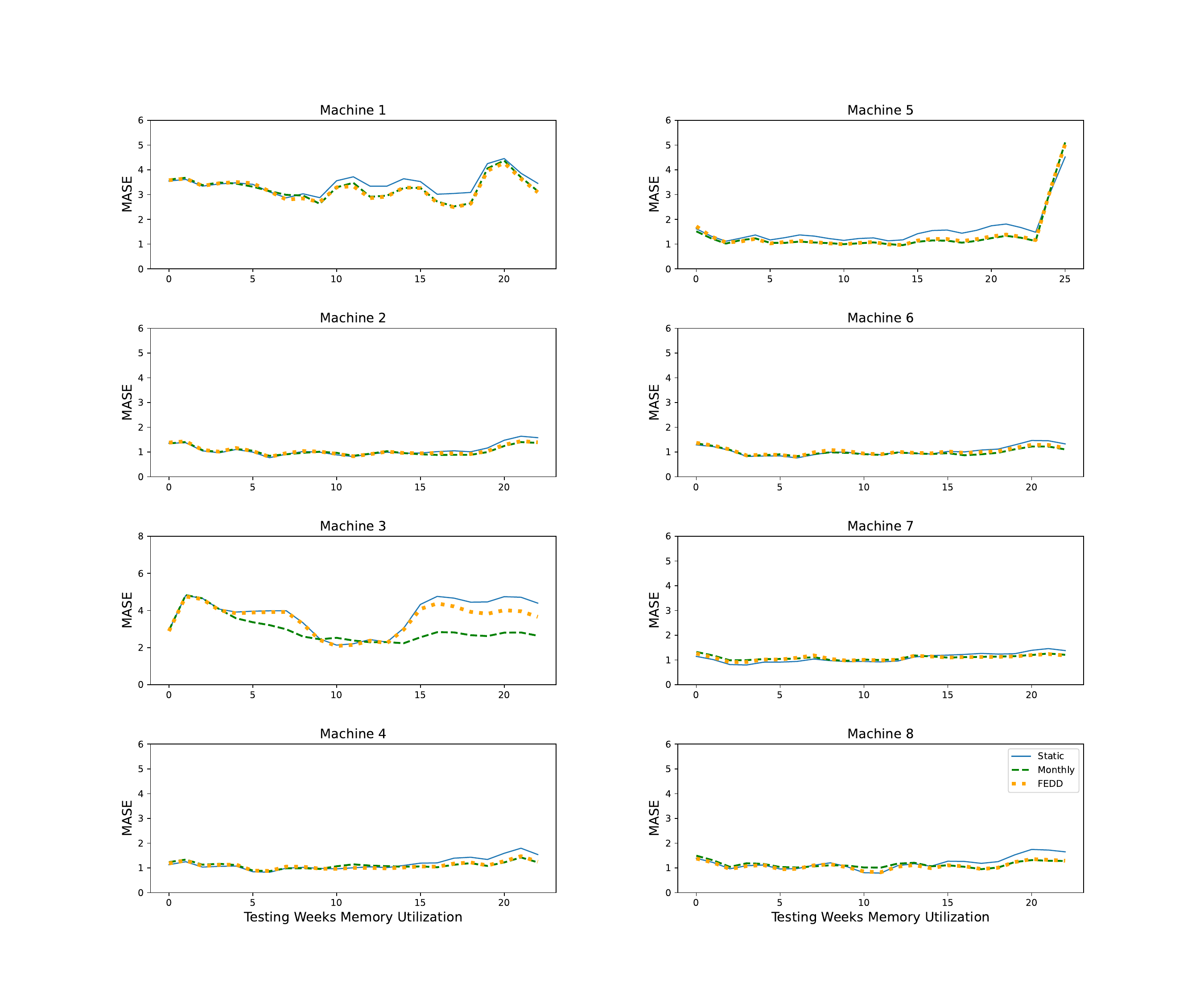}
    \caption{Memory Utilization}
    \label{figure:rq1_static_error_over_time_memory}
    \vspace{-1em}
\end{figure*}

In Fig.~\ref{figure:rq1_static_error_over_time_CPU} and Fig.~\ref{figure:rq1_static_error_over_time_memory}, we present the performance (MASE) of the forecasting model over time for CPU and memory utilization, respectively. This MASE is depicted in the two figures by the continuous, blue line called "static". As an overview of this experiment, we cannot observe a gradual performance degradation over time for all analyzed time series. In most situations, the model's performance is relatively constant with small fluctuations. These fluctuations can indicate that retraining is needed, but the changes do not severely impact the model's performance since, in most situations, the MASE variance is smaller than 1. When it comes to CPU utilization forecasting depicted in Fig.~\ref{figure:rq1_static_error_over_time_CPU}, the time series corresponding to Machine 3 suffers a severe increase in MASE in the first weeks. However, the MASE is drastically decreasing, indicating that the remaining time series becomes similar to the time series the model has been trained on. This suggests that the time series experiences a drastic change after a certain point and afterward changes back to its initial state. Thus, in this situation, especially, the model should be able to adapt to this temporary drift, which can be achieved by retraining the model. We can further notice in Fig.~\ref{figure:rq1_static_error_over_time_CPU} that in the case of Machine 5, the MASE is increasing between weeks 20-25, indicating that there is a change in the data around this period.

In Fig.~\ref{figure:rq1_static_error_over_time_memory}, we can notice that the performance of the forecasting model is gradually dropping since the MASE, which corresponds to the forecasting error, is increasing for Machines 2, 4, 6, 7, and 8. This might suggest that, for these cases, the model might benefit from being updated over time. When it comes to machines 1 and 3, it can be noticed that the drops in performance are not gradual, but rather sudden which can be observed by the fluctuations in the forecasting model's error in some specific time intervals, such as testing week 7 for Machine 3 and testing weeks 15 or 20 for Machine 1. These can be explained by fluctuations in the time series, which change suddenly and suddenly in different periods in a similar manner as in Machine 3 in the case of CPU utilization. Moreover, similar to the situation of CPU utilization, we can observe in Fig.~\ref{figure:rq1_static_error_over_time_memory} that Machine 5 also experiences a sudden decrease in performance (increase in MASE) between weeks 20-25.



\subsection{Periodic vs Drift Detection-based Retraining}

In this set of experiments, we answer RQ2 and understand what is the difference between our current way of retraining the forecasting model (monthly retraining) and retraining based on drift detection using FEDD. For this experiment, the model that is retrained on a monthly basis is our baseline, since we analyze the benefits of employing concept drift detection-based retraining over periodic retraining. To understand whether FEDD-based retraining is beneficial, we measure the improvement in forecast accuracy and the percentage of retraining saved while retraining based on FEDD compared to periodic retraining. We depict our results in Table~\ref{table:periodic_vs_FEDD}.

\begin{table}[ht!]
\centering
\caption{MASE Improvements (Impr.) and retraining savings in percentage between retraining periodically (1 week, 2 weeks, 1 month) and retraining using FEDD. The "+" symbol shows that the model retrained based on FEDD performed better than the periodic one and the "-" shows that it performed worse.}
\begin{tabular}{|c | c | r | r | r |r| r | r | r |r | r | r | r |}

 \cline{2-4}

  \multicolumn{1}{p{0.000000001cm}}{} &
 \multicolumn{1}{|c|}{\footnotesize{Time Series}} &
 \multicolumn{1}{|c|}{\footnotesize{MASE Impr. (\%)}} &
 \multicolumn{1}{|c|}{\footnotesize{Retraining Savings (\%)}} \\

 \cline{1-4}

 \multirow{8}{0.15em}{\rotatebox[origin=c]{90}{\scriptsize{\textbf{CPU Utilization}}}}& Machine 1 & +6.67 & 67 \\
 & Machine 2 & -2.06 & 67 \\
 & Machine 3  & -5.70 & 50  \\
 & Machine 4 & +0.00 & 50  \\
 & Machine 5 & -2.17 & 57 \\
 & Machine 6 & +1.14 & 50  \\
 & Machine 7 & -1.09 & 67  \\
 & Machine 8 & +2.22 & 50  \\

\hline
\hline

 \multirow{8}{0.15em}{\rotatebox[origin=c]{90}{\scriptsize{\textbf{Memory Utilization}}}}& Machine 1 & +1.52 & 50 \\
 & Machine 2 & -0.95 & 50  \\
 & Machine 3 & -19.20 & 67  \\
 & Machine 4 & +5.36 & 50  \\
 & Machine 5 & -3.73 & 57   \\
 & Machine 6 & -2.97 & 50  \\
 & Machine 7 & +4.42 & 67  \\
 & Machine 8 & +3.54 & 50  \\
 
\cline{1-4}
\end{tabular}

\label{table:periodic_vs_FEDD}

\end{table}

From Table~\ref{table:periodic_vs_FEDD} we can notice that in all situations, retraining based on drift detection implies 50\% savings in the number of required retrainings. This suggests that the forecasting model was retrained only half of the time using the chosen drift detector (FEDD) compared to monthly. When it comes to performance improvement, we can observe that the MASE obtained by drift detection-based retraining was, for half of the time series, lower than the one obtained by retraining periodically. The most dramatic decrease in performance was observed for the time series generated by Machine 3 for both CPU and memory utilization. In these cases, the MASE decreases by 5.7\% and 19.2\%, respectively, when the required retrainings decrease by 50\% and 67\%, respectively. This shows that lowering the number of retrainings comes with the high cost of lowering the MASE of the forecasting model. In the situation of time series generated by Machines 1, 4, and 8, the MASE generated by retraining based on drift detection resulted in an increase in the forecasting performance for both CPU and memory utilization. This shows that retraining based on drift detection is beneficial for both reducing the number of times the forecasting model requires retraining and obtaining better forecasting accuracy. 

Table~\ref{table:periodic_vs_FEDD} shows the MASE improvement when considering the average MASE on all testing batches for each retraining technique. In Fig.~\ref{figure:rq1_static_error_over_time_CPU} and Fig.~\ref{figure:rq1_static_error_over_time_memory}, we depict the MASE on each testing batch to better understand the MASE difference on each testing batch. The MASE corresponding to monthly retraining is depicted using a dashed line, while the MASE corresponding to retraining the forecasting model based on drift detection is depicted using a dotted line in the two figures. In both CPU and memory utilization, we do not observe significant differences in MASE between retraining monthly vs. retraining based on drift detection, except for the time series generated by Machine 3. For this particular machine, the forecasting model retrained based on drift detection performs similarly to the situation in which it was never retrained, but its MASE starts decreasing after testing batch 10 for CPU utilization (Fig.~\ref{figure:rq1_static_error_over_time_CPU}) and 15 for memory utilization~\ref{figure:rq1_static_error_over_time_memory}. However, for this situation, the MASE obtained by retraining periodically is, in most situations, substantially lower than the one obtained by retraining based on drift detection, suggesting that for this particular time series, retraining more often is beneficial. This can also suggest that the FEDD drift detector did not manage to capture all the situations in which retraining was required.


\section{Discussion}
Through this case study, we noticed that employing a drift detection-based retraining approach is beneficial in the context of capacity forecasting for our industry partner. Our conclusions are derived from the fact that we demonstrated how a drift detection-based retraining reduces the time required to update the forecasting models, while minimally impacting their forecasting performance. Furthermore, employing a drift detection-based retraining approach improves the scalability of the forecasting model to multiple time series. In this section, we are further discussing the implications of retraining the capacity forecasting model using drift detection. We evaluate these implications by examining the model's accuracy over time, highlighting scenarios where drift detection-based retraining underperformed compared to monthly retraining, and explaining the potential reasons for this. Additionally, we explore the design considerations for machine learning systems when implementing drift detection-based retraining, emphasizing key factors that machine learning engineers must take into account.

\subsection{Accuracy Implications of Retraining based on Drift Detection}

\begin{figure*}[ht] 
\centering 
\subfloat[]{%
\includegraphics[width=0.95\textwidth]{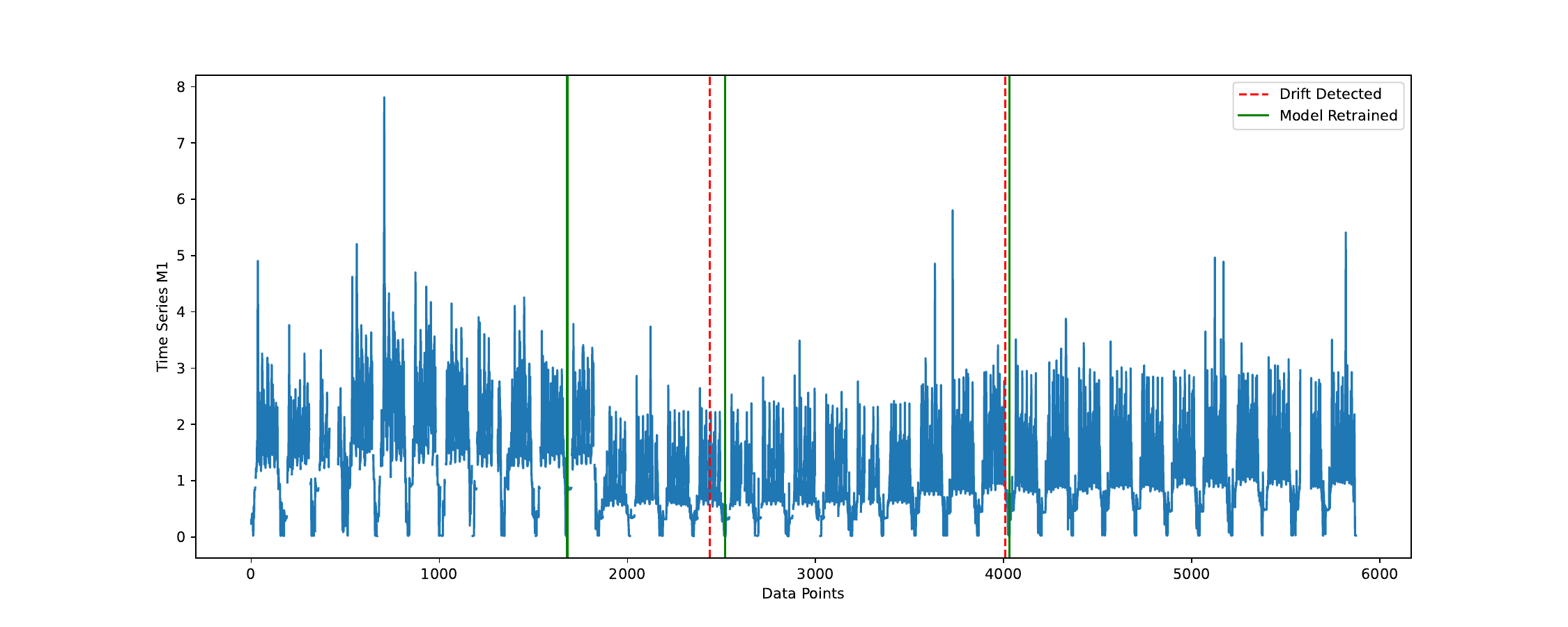} \label{fig:machineM1} } \hfill \subfloat[]{
\includegraphics[width=0.95\textwidth]{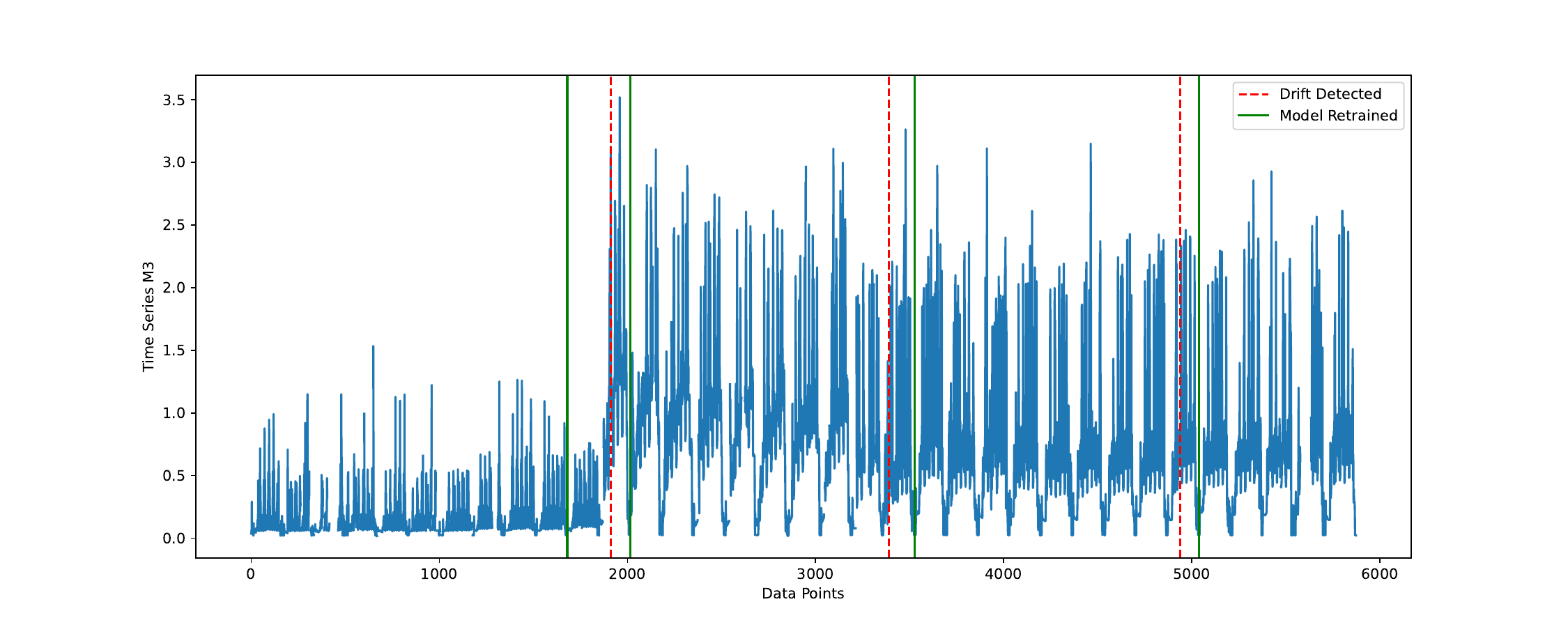} \label{fig:MachineM3} } \caption{Time Series Corresponding to CPU utilization for Machine 1 (a) and Machine 3 (b) including the moments when the drift was detected by FEDD and the moments when the forecasting model was retrained.} \label{fig:combined_discussion} 
\end{figure*}

From our experiments, we observed that in most of the situations, the performance of the forecasting model is not highly impacted when retraining based on drift detection using FEDD compared to our current retraining practice, monthly retraining. In some situations, we observed even an improvement in forecasting accuracy when retraining based on drift detection. This shows that retraining when FEDD identifies changes in the time series data is a promising solution to reduce the number of times the forecasting model requires retraining, while not significantly impacting the model's accuracy. However, this conclusion does not hold when it comes to the time series generated by Machine 3 as observed in both Fig.~\ref{figure:rq1_static_error_over_time_CPU} and Fig.~\ref{figure:rq1_static_error_over_time_memory}. In this specific case, we observe that the model that is periodically retrained performs significantly better than the model retrained based on drift detection. Thus, we performed an in-depth analysis of this case, and we present our findings in this section. We are using Fig.~\ref{fig:combined_discussion} to explain our findings.

One interesting finding from our analysis is that while all selected time series generated by all machines experience changes in the data, the time series generated by Machine 3 are different in terms of how frequently they experience data changes and how long the changes last. Thus, a major characteristic of the time series generated by Machine 3 is the short and sudden changes in the data, as can also be seen in Fig.~\ref{fig:MachineM3}.

We investigated whether the type of data change, a sudden change in the time series or a more gradual change in the time series, is a limitation of the FEDD drift detector. However, from Fig.~\ref{fig:machineM1} we can notice that FEDD managed to identify both a sudden data change (first identified drift) and a more gradual data change (second identified drift), in this example corresponding to Machine 1. We observed similar behavior in other time series but only decided to show Machine 1 since the difference between the data changes can be visibly noticed. Thus, the type of data change is not a limitation of FEDD, but the fact that the changes in the time series are short and might be related to the way FEDD was designed. 

Once it detects a drift, FEDD enters a period of inactivity until it gathers enough data to start detecting again. During this period, according to the detector's functionality~\cite{fedd}, FEDD shifts its reference data with a predefined window size to avoid continuously triggering false alarms. This period of inactivity can be the main reason why FEDD did not find any drift between data points 2000 and 3000 from Fig~\ref{fig:MachineM3}, although the data changed significantly. Also, the first model retraining around data point 2000 did not capture enough changed data to allow the forecasting model to learn the new time series pattern, which can explain the low performance. This argument is further supported by the fact that the forecasting model retrained periodically achieved significantly better performance than the one trained based on drift detection. Therefore, our findings suggest that FEDD is a suitable solution as long as the changes in the time series are not frequent. For these types of time series, periodic retraining is the most suitable solution to preserve the forecast performance. Therefore, we consider building a hybrid retraining approach, in which time series with short sudden changes (e.g. the ones generated by Machine 3) are retrained periodically, while the others are retrained based on drift detection. This hybrid approach will allow one to reduce the retraining time, while ensuring that the forecast performance is not impacted.




\subsection{Forecasting System Design Implications}

When it comes to designing a drift detection-based retraining pipeline for a time series forecasting model, we consider FEDD a suitable solution when scalability is a major requirement. The reason for this is that FEDD does not require storing the entire time series, and continuously accessing it to detect drift, which could come with significant storage expenses and latency when accessing the data. 

A major concern that practitioners need to consider is handling missing data, as FEDD is not designed to handle missing samples. Although in our solution we opted to remove them, this might have implications for the way FEDD detects drift. For instance, a higher number of missing data points could result in a distorted time series, which FEDD could see as a drift and erroneously signal the need to retrain the model. Although in our experiments, we did not encounter this situation, we recommend practitioners be aware of this limitation of FEDD when applying it to real-world data. To avoid false alarms generated by missing samples, we suggest integrating FEDD with a missing values monitoring block to better investigate whether the drift was signaled due to a high number of missing data points. Furthermore, we recommend that researchers incorporate missing data handling techniques in time series drift detection and assess their implications in drift detection accuracy.

In terms of the moment when the forecasting model is updated, we did not consider retraining the model immediately after a drift was detected, since it is unsuitable in practical scenarios. The reason for this is the availability of a responsible data scientist to ensure that, after retraining, the model can be securely and safely deployed. For instance, if a drift is detected outside of working hours, the data scientist is not available to perform model quality checks after it has been retrained and redeployed.


\section{Threats to Validity}
In this case study, we focus mainly on the forecasting of capacity for CPU and memory resources using data from ING. However, this use case offers a unique and realistic setting, a large-scale financial infrastructure where time forecasting models are continuously exposed to evolving patterns in the data. Therefore, the results highlight the effectiveness of the process presented. Incorporating concept drift detection-based retraining in forecasting pipelines offers valuable insight for practitioners navigating similar contrasts in high-stakes environments.

\section{Conclusions}

This work investigates the impact of retraining a time series forecasting model for capacity management based on drift detection. To identify drift and, therefore, the need to retrain the forecasting model, we employ the FEDD drift detector. In our experiments, we investigate the effect of retraining based on FEDD compared to retraining monthly in terms of the performance of the forecasting model and how many times the model requires retraining. The goal of this study is to understand whether FEDD can be used to reduce the number of times the forecasting model requires retraining, since with time, if the model is applied to more time series, we might encounter scalability issues that will no longer allow monthly model updates. 

Our experiments suggest that for both CPU and memory utilization-related time series, retraining based on drift detection does not imply a significant drop in the forecasting model's performance. The only situation in which we observed that the periodically retrained model predicts significantly better than the one retrained based on FEDD is for the time series that experience short and sudden changes. For these time series, particularly, we further suggest employing a periodic retraining approach. Furthermore, in this paper, we discuss the implications of employing FEDD in a practical scenario and the challenges that we encountered in terms of integration with the forecasting model. For instance, we highlighted the fact that FEDD cannot handle missing values and should be carefully considered and evaluated in situations in which the number of missing samples in a time series is significantly high.

Future work should focus on improving FEDD to work with distorted time series due to missing values. Furthermore, one major limitation of FEDD that we discovered is handling short, sudden changes in the time series due to the period of restart that FEDD requires after detecting a drift. For this reason, we consider that developing a drift detector with a shorter restart period, which does not signal a high number of false alarms, is required.


\bibliographystyle{splncs04}
\bibliography{bibliography}

\begin{thebibliography}{10}
\providecommand{\url}[1]{\texttt{#1}}
\providecommand{\urlprefix}{URL }
\providecommand{\doi}[1]{https://doi.org/#1}

\bibitem{eddm}
Baena-García, M., Campo-Ávila, J., Fidalgo-Merino, R., Bifet, A., Gavald, R., Morales-Bueno, R.: Early drift detection method (2006)

\bibitem{incident1}
Batta, R., Shwartz, L., Nidd, M., Azad, A.P., Kumar, H.: A system for proactive risk assessment of application changes in cloud operations. In: 2021 IEEE 14th International Conference on Cloud Computing (CLOUD). pp. 112--123 (2021). \doi{10.1109/CLOUD53861.2021.00025}

\bibitem{BAYRAM2022108632}
Bayram, F., Ahmed, B.S., Kassler, A.: From concept drift to model degradation: An overview on performance-aware drift detectors. Knowledge-Based Systems  \textbf{245},  108632 (2022)

\bibitem{adwin}
Bifet, A., Gavaldà, R.: Learning from time-changing data with adaptive windowing. In: SDM. vol.~7 (2007)

\bibitem{diskfailureprevwork1}
Botezatu, M.M., Giurgiu, I., Bogojeska, J., Wiesmann, D.: Predicting disk replacement towards reliable data centers. In: ACM KDD (2016)

\bibitem{fedd}
Cavalcante, R.C., Minku, L.L., Oliveira, A.L.I.: Fedd: Feature extraction for explicit concept drift detection in time series. In: 2016 International Joint Conference on Neural Networks (IJCNN). pp. 740--747 (2016)

\bibitem{usingtsdriftdetectioninrealapplications}
Cavalcante, R.C., Oliveira, A.L.I.: An approach to handle concept drift in financial time series based on extreme learning machines and explicit drift detection. In: 2015 International Joint Conference on Neural Networks (IJCNN). pp.~1--8 (2015). \doi{10.1109/IJCNN.2015.7280721}

\bibitem{salesforceai}
{Cheng}, Q., {Sahoo}, D., {Saha}, A., {Yang}, W., {Liu}, C., {Woo}, G., {Singh}, M., {Saverese}, S., {Hoi}, S.C.H.: {AI for IT Operations (AIOps) on Cloud Platforms: Reviews, Opportunities and Challenges}. arXiv e-prints arXiv:2304.04661 (Apr 2023). \doi{10.48550/arXiv.2304.04661}

\bibitem{aiopschallenges}
Dang, Y., Lin, Q., Huang, P.: Aiops: Real-world challenges and research innovations. In: IEEE/ACM ICSE-Companion. pp.~4--5 (2019)

\bibitem{efte}
Ding, F., Luo, C.: The entropy-based time domain feature extraction for online concept drift detection. Entropy  \textbf{21}(12) (2019). \doi{10.3390/e21121187}, \url{https://www.mdpi.com/1099-4300/21/12/1187}

\bibitem{googletraceprediction1}
El-Sayed, N., Zhu, H., Schroeder, B.: Learning from failure across multiple clusters: A trace-driven approach to understanding, predicting, and mitigating job terminations. In: IEEE ICDCS. pp. 1333--1344 (2017)

\bibitem{conceptdriftadaptation}
Gama, J.a., \v{Z}liobaitundefined, I., Bifet, A., Pechenizkiy, M., Bouchachia, A.: A survey on concept drift adaptation. ACM Comput. Surv.  \textbf{46}(4) (2014)

\bibitem{ddm}
Gama, J., Medas, P., Castillo, G., Rodrigues, P.: Learning with drift detection. In: SBIA. vol.~8, pp. 286--295 (2004)

\bibitem{incident2}
Güven, S., Murthy, K., Shwartz, L., Paradkar, A.: Towards establishing causality between change and incident. In: NOMS 2016 - 2016 IEEE/IFIP Network Operations and Management Symposium. pp. 937--942 (2016). \doi{10.1109/NOMS.2016.7502929}

\bibitem{feddPSO}
H.~F. M.~Oliveira, G., C.~Cavalcante, R., G.~Cabral, G., L.~Minku, L., L.~I.~Oliveira, A.: Time series forecasting in the presence of concept drift: A pso-based approach. In: 2017 IEEE 29th International Conference on Tools with Artificial Intelligence (ICTAI). pp. 239--246 (2017). \doi{10.1109/ICTAI.2017.00046}

\bibitem{forecastevaluation}
Hewamalage, H., Ackermann, K., Bergmeir, C.: Forecast evaluation for data scientists: common pitfalls and best practices. Data Mining and Knowledge Discovery  \textbf{37},  788 -- 832 (2022), \url{https://api.semanticscholar.org/CorpusID:247594079}

\bibitem{masemetric}
Hyndman, R.J., Koehler, A.B.: Another look at measures of forecast accuracy. International Journal of Forecasting  \textbf{22}(4),  679--688 (2006). \doi{https://doi.org/10.1016/j.ijforecast.2006.03.001}, \url{https://www.sciencedirect.com/science/article/pii/S0169207006000239}

\bibitem{ericsonanomalydetection}
Isaac, E.R.H.P., Sharma, A.: Adaptive thresholding heuristic for kpi anomaly detection (2023), \url{https://arxiv.org/abs/2308.10504}

\bibitem{Isaac2023QBSDQS}
Isaac, E.R.H.P., Singh, B.: Qbsd: Quartile-based seasonality decomposition for cost-effective time series forecasting (2023), \url{https://api.semanticscholar.org/CorpusID:260926748}

\bibitem{eileenpaper}
Kapel, E., Cruz, L., Spinellis, D., Van~Deursen, A.: On the difficulty of identifying incident-inducing changes. In: Proceedings of the 46th International Conference on Software Engineering: Software Engineering in Practice. p. 36–46. ICSE-SEIP '24, Association for Computing Machinery, New York, NY, USA (2024). \doi{10.1145/3639477.3639755}, \url{https://doi-org.tudelft.idm.oclc.org/10.1145/3639477.3639755}

\bibitem{missingdataperspective}
Li, S.C.X., Marlin, B.M.: Learning from irregularly-sampled time series: a missing data perspective. In: Proceedings of the 37th International Conference on Machine Learning. ICML'20, JMLR.org (2020)

\bibitem{nodefailurepredictionconceptdrift}
Li, Y., Jiang, Z.M.J., Li, H., Hassan, A.E., He, C., Huang, R., Zeng, Z., Wang, M., Chen, P.: Predicting node failures in an ultra-large-scale cloud computing platform: An aiops solution. ACM Trans. Softw. Eng. Methodol.  \textbf{29}(2) (2020)

\bibitem{modelmaturity}
Lyu, Y., Li, H., Jiang, Z.M., Hassan, A.E.: Assessing the maturity of model maintenance techniques for aiops solutions. arXiv preprint arXiv:2311.03213  (2023)

\bibitem{datasplittingdecisions}
Lyu, Y., Li, H., Sayagh, M., Jiang, Z.M.J., Hassan, A.E.: An empirical study of the impact of data splitting decisions on the performance of aiops solutions. ACM Trans. Softw. Eng. Methodol.  \textbf{30}(4) (jul 2021)

\bibitem{towardsaconsistentinterpretation}
Lyu, Y., Rajbahadur, G.K., Lin, D., Chen, B., Jiang, Z.M.J.: Towards a consistent interpretation of aiops models. ACM Trans. Softw. Eng. Methodol.  \textbf{31}(1) (2021)

\bibitem{diskfailureprevwork}
Mahdisoltani, F., Stefanovici, I., Schroeder, B.: Proactive error prediction to improve storage system reliability. In: 2017 USENIX Annual Technical Conference (USENIX ATC 17). pp. 391--402 (2017)

\bibitem{rootcauseanalysis1}
Nguyen, H., Tan, Y., Gu, X.: Pal: Propagation-aware anomaly localization for cloud hosted distributed applications. In: Managing Large-Scale Systems via the Analysis of System Logs and the Application of Machine Learning Techniques. SLAML '11, Association for Computing Machinery, New York, NY, USA (2011). \doi{10.1145/2038633.2038634}, \url{https://doi.org/10.1145/2038633.2038634}

\bibitem{myshortpaper}
Poenaru-Olaru, L., Cruz, L., Rellermeyer, J., {Van Deursen}, A.: Maintaining and monitoring aiops models against concept drift. In: Proceedings - 2023 IEEE/ACM 2nd International Conference on AI Engineering - Software Engineering for AI, CAIN 2023. pp. 98--99. Proceedings - 2023 IEEE/ACM 2nd International Conference on AI Engineering - Software Engineering for AI, CAIN 2023, Institute of Electrical and Electronics Engineers (IEEE), United States (2023). \doi{10.1109/CAIN58948.2023.00024}

\bibitem{mypaperanomalydetection}
Poenaru-Olaru, L., Cruz, L., Rellermeyer, J.S., van Deursen, A.: Is your anomaly detector ready for change? adapting aiops solutions to the real world. In: CAIN24, 2nd International Conference on AI Engineering - Software Engineering for AI (2024)

\bibitem{rootcauseanalysis2}
Shan, H., Chen, Y., Liu, H., Zhang, Y., Xiao, X., He, X., Li, M., Ding, W.: ?-diagnosis: Unsupervised and real-time diagnosis of small- window long-tail latency in large-scale microservice platforms. In: The World Wide Web Conference. p. 3215–3222. WWW '19, Association for Computing Machinery, New York, NY, USA (2019). \doi{10.1145/3308558.3313653}, \url{https://doi-org.tudelft.idm.oclc.org/10.1145/3308558.3313653}

\bibitem{prophet}
Taylor, S., Letham, B.: Forecasting at scale. The American Statistician  \textbf{72} (09 2017). \doi{10.1080/00031305.2017.1380080}

\bibitem{Wang2022ANT}
Wang, K., Tan, Y., Zhang, L., Chen, Z., Lei, J.: A network traffic prediction method for aiops based on tda and attention gru. Applied Sciences  (2022), \url{https://api.semanticscholar.org/CorpusID:253034811}

\bibitem{nodefailure1}
Xu, Y., Sui, K., Yao, R., Zhang, H., Lin, Q., Dang, Y., Li, P., Jiang, K., Zhang, W., Lou, J.G., Chintalapati, M., Zhang, D.: Improving service availability of cloud systems by predicting disk error. In: Proceedings of the 2018 USENIX Conference on Usenix Annual Technical Conference. p. 481–493. USENIX ATC '18 (2018)

\end{thebibliography}

\end{document}